\documentclass[12pt,letterpaper]{article}
\usepackage{amsmath,amssymb,pgf,pgfarrows,pgfnodes,float,appendix, hyperref}
\usepackage{graphicx}
\usepackage{subfigure}
\usepackage[margin=0.9in]{geometry}

\newcommand{\be}{\begin{equation}}
\newcommand{\ee}{\end{equation}}
\newcommand{\bea}{\begin{eqnarray}}
\newcommand{\eea}{\end{eqnarray}}

\title{{\rm\footnotesize \qquad \qquad \qquad \qquad \qquad \ \qquad \qquad \qquad \ \ \ \ \ \                  RUNHETC-2020-33}\vskip.5in   Microscopic Models of Linear Dilaton Gravity and Their Semi-classical Approximations}
\author{Tom Banks\\
Department of Physics and NHETC\\
Rutgers University, Piscataway, NJ 08854\\
E-mail: \href{mailto:tibanks@ucsc.edu}{tibanks@ucsc.edu}
\\
\\
}
\date{}
\begin{document}
\maketitle

\begin{abstract} We reanalyze and expand upon models\cite{tb1+1} for linear dilaton black holes, and use them to test several ideas about black hole physics that have been proposed in the context of the AdS/CFT correspondence.  In particular, we examine ideas based on the definition of quantum extremal surfaces in quantum field theory in curved space-time\cite{qes}.  The low energy effective field theory of our model is the large $N$ CGHS\cite{cghs} model. The leading order large $N$ solution of that model includes the one loop effects that are taken into account in the "island" proposal for understanding the Page curve.  Contrary to the results of the island analysis, that solution leads to a singular geometry for the evaporated black hole.  If the singularity obeys Cosmic Censorship then Hawking evaporation leaves behind a remnant object with a finite fraction of the black hole entropy. If the singularity becomes naked at some point, boundary conditions on a time-like line emanating from that point can produce a sensible model where we expect a Page curve\cite{rst3}.  We show that the fully UV complete model gives a correct Page curve, as it must since the model is manifestly unitary.  However the correct computation of the entanglement entropy of the outgoing radiation in the black hole background requires one to understand the ground state entanglement entropy of the fermion fields defining the model, with full non-perturbative precision.  The vacuum subtraction depends on the microscopic details of the many different models that lead to the same semiclassical black hole physics.  Together, these results suggest that one cannot understand completely the unitarity of black hole evaporation within the formalism of perturbative quantum field theory in curved space-time.  Recent results\cite{replicawormholes} suggest that the island formula, which appears to involve only one loop computations, in fact encodes non-perturbative contributions to the gravitational path integral. The question of {\it why} Euclidean gravity computations can capture information about microscopic states of quantum gravity remains mysterious. In a speculative coda to the paper we suggest that the proper way of understanding the relation between space-time geometry and entropy is via Jacobson's\cite{ted} interpretation of general relativistic field equations as the hydrodynamic equations of the area law for the maximal entropy of causal diamonds.  The hydrodynamic equations of any quantum system are stochastic classical equations and we suggest that the remarkable results of\cite{sss} might be understood in the context of the functional integral solution of these stochastic equations.  

\end{abstract}

\section{Introduction}

There has been a lot of activity recently, devoted to the SYK\cite{syk} model and its connection to a model of $1 + 1$ dimensional classical gravity, the JT model\cite{jt}\footnote{More properly, in light of recent discoveries, the JB model.} .   The purpose of the present paper is to review and expand upon work done\cite{tb1+1} on another model of $1 + 1$ dimensional quantum gravity, the linear dilaton model\cite{cghs}, or more properly the linear dilaton model coupled to $N$ massless fermion fields.   This was our first example of what today would be called a near horizon limit.  It is the near horizon limit of charged black holes in Type II string theory compactified on a Calabi-Yau manifold.  These black holes are extremal, but not BPS.   The fermion fields arise as limits in the black hole throat of Ramond-Ramond states, whose number depends in an intricate way on the topology of the $CY_3$.  It is reasonable to imagine that $N$ can be very large, though probably not infinite.  It's also reasonable to imagine that there are other kinds of perturbative string models whose low energy effective field theory has similar linear dilaton black holes with the same near horizon limit.   The models are exactly integrable at the classical level (after bosonization of the fermions) and provide examples of renormalizable field theories of quantum gravity.  They contain a single dimensionful parameter, and a dimensionless parameter, the value of the dilaton field at infinity.  We know quite a lot about quantum theories that give rise to these low energy approximations, so it is abundantly clear that quantization of the model as a field theory does not give the correct answers.

Indeed, quite remarkably, the case $N = 2$ is the low energy effective theory, at strong string coupling, of one of our basic examples of an exactly soluble string theory.  Black hole solutions of the low energy field theory of the weakly coupled model were found in\cite{1+1bh}, but the quantum theory has no black holes\cite{maldakstrom}. We will not go through the extraordinary history\cite{oldmm} of the $N = 2$ model, but will remark only that the basic quantum mechanical model was solved in 1977.  A certain limit of it was recognized to be a string theory in 1990, and the non-perturbative solution was completely understood only in 2003.   We will rely mainly on the results of \cite{moore} and \cite{akk}.  The reader who is not acquainted with the details of these models can find a short summary in Appendix C.  It should be consulted in order to understand the results of the current paper.

The outline of the results of these investigations is

\begin{itemize}

\item The proper quantum description the model starts with a non-relativistic model of non-interacting fermions with single particle Hamiltonian $(p^2 - \lambda^2) + \mu$. String perturbation theory is the large $\mu$ expansion.  From the string theory point of view, large $\mu$ is a repulsive potential for the Weyl fermions, which prevents them from exploring the region in the vacuum solution where the string coupling is strong.
The models of\cite{tb1+1} add interactions localized near the top of the inverted oscillator, which are invisible in string perturbation theory.  Although, like the Dirac operator, the inverted oscillator Hamiltonian is unbounded from below, Moore\cite{moore} showed that the field theory is perfectly well defined and has scattering states at $\lambda = \pm\infty$.  Moore begins with a large $\lambda$ cutoff that makes the Fermi sea finite (interestingly, this is the first example of a UV/IR connection in the history of string theory), and then shows that there is a cutoff independent scattering matrix.
The work of\cite{akk} shows that these states are a pair of Weyl fermions propagating in a Minkowski space derived from the underlying model.  The analysis of\cite{akk} is rather simple and illuminating.  Going to the variables $x_{\pm} = \frac{\lambda \pm p}{\sqrt{2}} $ one finds that the model has a scattering matrix, where incoming/outgoing states are Weyl fermion fields $\psi_W^{\pm}  = \psi ({\rm ln}\ x_{\pm} ) $.   These are peculiar field operators in the non-relativistic Hilbert space, but perfectly well defined.  The scattering matrix for $\mu = 0$ has exactly the form conjectured by 't Hooft for the black hole S matrix.

\item The model has a perturbation series identical to (Type 0B) string theory, but the perturbation theory is not Borel summable\cite{shenker}, an ambiguity which we will understand more deeply, and exploit, in the present paper.  Perturbatively there are two decoupled fermion modes, but the model that keeps only one of them is not well defined.  The authors of\cite{mooreand?} recognized that an additional ambiguity was the ability to add a potential confined to the very top of the well.  Perturbative scattering captures only the way in which fermions bounce off the potential wall at large $\lambda$, which exists when $\mu$ is large.  As in higher dimensional theories, effects that might be associated with black holes appear, at weak string coupling, only above a very high energy threshold.

\item The gravitational field plays absolutely no role in the quantum theory.  The space of solutions of the pure gravity model consists only of linear dilaton black holes (and the value of the generator of the asymptotic time translation symmetry).  The bosonized $N = 2$ model has classical solutions in which black holes are created by scattering coherent states of the fermion density operator, but rigorous analysis has shown that there are no such black hole excitations of the model.  This is true both at the level of non-relativistic bosonization\cite{polchinski} and via a computation of the exact S-matrix\cite{moore}\cite{mooreand?}\cite{akk}.   Note that\cite{polchinski2} claimed to see a contribution of the graviton field to the "leg poles" of the string amplitudes.  These are an example of the extreme CDD ambiguity\cite{zamo} of massless scattering in $1 + 1$ dimensions.  They do not appear in the fermionic computation of the amplitudes and the authors of\cite{yinetal} have argued that they can be eliminated by a "proper" normalization of string vertex operators.  The authors of\cite{cghs} argued that the classical computation of black hole formation could not be trusted, because the semiclassical expansion parameter, the string coupling, was not small at the horizon of the putative black hole. They showed that taking $N$ large ameliorated this problem.   However, as we shall see, large $N$ is not sufficient.

\item CGHS were probably the first authors to point out that the value of the field multiplying the Einstein term in scalar tensor theories of gravity in two dimensions should be thought of as the standin for the non-existant transverse area of a horizon. This field counts entropy.

\item The entropy counting formula can be used to "derive" the gravitational picture from the underlying fermionic model.  Using Thomas-Fermi methods to count fermionic entropy along the $\lambda$ axis one finds the mapping between $\lambda$ and $x$ which converts that formula into the static dilaton profile of the classical gravito-dilaton background\cite{dasetal} for $\mu = 0$. For large $\mu$ (weak string coupling), one must go to larger $\lambda$ before the classical gravitational entropy profile is valid.

\item The analysis of\cite{akk} shows that the exact S-matrix has the 't Hooft-Dray form\cite{thooft}
\begin{equation} S = S_{in} S_{x_{\pm}} S_{out} , \end{equation} where $S_{x_{\pm}}$ is the Fourier transform operator between in and outgoing near horizon null coordinates, assumed to satisfy 
canonical commutation relations.  In the fermion model,  $x_{\pm} = \frac{\lambda \pm p}{\sqrt{2}} $.  $S_{in/out}$ are simply the aforementioned transformation between these variables and the classical space-time coordinates.  Note that the use of the $x_{\pm}$ variables makes it immediately obvious that the scattering states are relativistic Weyl fermions.  This connection to the 't Hooft commutation relations was originally noted by\cite{verlinde}.  We should point out that these relations were identified by Shenker and Stanford\cite{sspolch} as avatars of maximal quantum chaos\cite{mssberry}.  Note however that although this form of the S-matrix was originally posited on the basis of black hole physics and high energy gravitational scattering, it exists in a completely integrable model, with no black hole excitations.

\item In\cite{tb1+1} we put all these pieces together to understand how the large $N$ model could be modified to produce quantum excitations with all the qualitative properties of linear dilaton black holes.  

\end{itemize}

The purpose of the present paper is to extend the analysis of\cite{tb1+1}, and to describe how certain speculative ideas that appear in the literature fare in the context of this model.  
The first set of ideas that is in tension with the model of this paper are those based on quantum corrected Ryu-Takyanagi formulae\cite{qes} and other attempts to understand black hole entropy and the Page curve in terms of leading order corrections to the semiclassical approximation, using quantum field theory in curved space-time.  In the models studied in this paper, the existence of black holes, and all questions about their properties are invisible at all orders of the weak coupling expansion.  In the context of AdS/CFT correspondences with a weak coupling string expansion,  the quantum RT formulae can be thought of as coming at leading order in the $\alpha^{\prime}$ expansion and one loop in the string coupling expansion, applied to a thermal state above the Hawking-Page transition.  In order to address issues of the Page curve in black hole evaporation, the holographic CFT must be coupled to an auxiliary system, to destabilize the thermal ensemble. The models in this paper have genuine meta-stable black holes.

We will show explicitly that these models have a Page curve, and that by going far out along the $\lambda$ axis the computation of the entanglement entropy of the Hawking radiation agrees with that in low energy field theory, modulo some possible issues with the proper subtraction of the vacuum entanglement entropy.  This validates the computation of the Page curve\cite{hss} in models obeying the boundary condition of\cite{rst3}.  However, the computation in\cite{hss} does not actually use that boundary condition, so that we need further evidence, within the semi-classical approach, that the so-called thunderpop boundary condition is implied by the derivation of the island formula.

The second AdS/CFT derived idea whose utility is called into question by our models is the $ER = EPR$ hypothesis\cite{erepr}: that entanglement of two holographic quantum systems is equivalent to some sort of Einstein-Rosen bridge between geometries.  This has been extended to the construction of "tranversable wormholes", corresponding to coupling two holographic quantum systems to each other\cite{traversable}.  In our model, we couple together $N$ copies of an exactly soluble string theory, each of which is dual to a string scale two dimensional geometry.  
The exactly soluble models have no black hole excitations.  The space-time interpretation of the coupled model is a single copy of the original geometry, which does have semi-classical black hole excitations.   The geometrical fields appear nowhere in the underlying quantum mechanics, but rather arise as hydrodynamic variables, describing entropy and energy flow.  One might attribute this failure of the ER=EPR paradigm to the fact the original space-times did not have "macroscopic curvature", but we can repeat the same exercise by coupling together two large $N$ models, each of which contains macroscopic black holes, to make a model with $N \rightarrow 2N$ which again has only a single macroscopic spacetime interpretation.  Of course, one can always claim that the single space-time in these models is really multiple copies of the space-time glued together by Planck scale wormholes, but that description has no utility for understanding the physics.

Finally, we will present a conjecture, which attempts to explain the success of\cite{sss} in calculating coarse grained properties of the energy spectrum of the SYK model in terms of the Euclidean path integral of JT gravity, summed over topologies.  Following Jacobson\cite{ted} our interpretation of classical gravitational equations as hydrodynamics leads one to consider the statistical form of modern hydrodynamics, used extensively in the study of turbulence\cite{stathydro}.  Adding a random stirring force to the Navier-Stokes equation leads to a functional integral approach to hydrodynamics, replete with Feynman diagrams and all of the usual apparatus of quantum field theory\cite{stathydro}.  This approach reveals "long time tail" corrections to classical hydrodynamic predictions, which have been observed in numerical experiments\cite{expt}.  Recent derivations of hydrodynamic equations for quantum systems\cite{hongmukundtbal} show that the statistical form of the hydrodynamic equations has its origin in quantum mechanics.  Indeed these explicit results were anticipated by the quantum fluctuation dissipation theorem\cite{fld}.  We propose to interpret the results of\cite{sss} as the proper statistical hydrodynamics of the SYK model.

\section{The Models}

We consider $N$ non-relativistic fermion fields with Hamiltonian
\begin{equation} H = \int d\lambda\  \psi_a^{\dagger} (\lambda) [\frac{1}{2} (p^2 - \lambda^2) + \mu] \psi_a (\lambda ) +  H_{int}\end{equation}\begin{equation}H_{int} = \int d\lambda d\kappa\ K(\lambda, \kappa) [- \frac{g^2}{N} \psi_a^{\dagger} (\lambda) \psi_a (\lambda) \psi_b^{\dagger} (\kappa) \psi_b (\kappa)  + \frac{J_{abcd}}{N^{3/2}} \psi_a^{\dagger} (\lambda) \psi_b (\lambda) 
\psi_c^{\dagger} (\kappa) \psi_d (\kappa) ] . \end{equation}  Here $J_{abcd}$ are chosen from independent but identical Gaussian distributions with mean $0$ and covariance that is order $1$ in the large $N$ limit.  $g^2$ is also order one.  $K(\lambda , \kappa)  = \langle \lambda | f(\Omega) | \kappa \rangle , $ where $\Omega  = p^2 + \lambda^2$ and $f(\Omega )$ is a smooth function vanishing at infinity faster than any inverse power of $\Omega $.  A simple example is $f = e^{-\beta \Omega}$. All the parameters in $f$ are kept fixed in the large $N$ limit, in units of the spring constant of the inverse oscillator.   The function $f$ has the property that the matrix elements of the interaction term between coherent states centered far from $p = \lambda = 0$ in classical single particle phase space are exponentially small.  Thus, scattering states with energy far below the top of the barrier are insensitive to the interaction.

This model has a "soluble" large $N$ limit (Appendix B), but we will avoid exploiting that because it introduces many conservation laws that are violated at any finite $N$.  Note also that at large $N$, the $J_{abcd}$ term includes a correction to $g^2$ that is down by $1/N$.
We choose the parameters so that, even at finite $N$ the correction does not change the qualitative physics.

Let us first set $J_{abcd} = 0$.   If the fermi surface of the non-interacting fermions is taken far below the top of the potential, then there is a systematic power series expansion of scattering amplitudes in inverse powers of $\mu$\cite{oldmm} .  Those perturbative amplitudes are $2N$ non-interacting copies of the amplitudes of the Type 0B string.
If we use logic often invoked in discussions of the AdS/CFT correspondence, we might consider these to be different copies of a $1 + 1$ dimensional universe, with any interaction between them being some kind of wormhole.   In fact, the space-time interpretation of these models is quite different.

The expectation value of the Hamiltonian in states with the expectation value of the density operator $n(\lambda) \equiv \frac{1}{N}\psi_a (\lambda) \psi_a (\lambda) = n_s (\lambda )$ is, to leading order in $1/N$ and using large $N$ factorization,
\begin{equation} \langle H \rangle =N[ \langle \int\ \psi_a^{\dagger} (- \partial_{\lambda}^2 - \lambda^2 - \mu )   \psi_a (\lambda) \rangle +  \int d\lambda d\kappa\ K(\lambda, \kappa) [- g^2 n_0 (\lambda ) n_0 (\kappa )] .\end{equation}  This is $N$ times the ground state energy of a single particle in the self consistent potential:
 \begin{equation} U(\lambda ) = - \lambda^2 - \int d\kappa K(\lambda , \kappa ) n_0 (\kappa ) .  \end{equation}  The self consistency equation is the requirement that $n_0$ be the electron number density expectation value in that ground state.
 
\begin{figure}[h!]
\begin{center}
 \includegraphics[width=12cm]{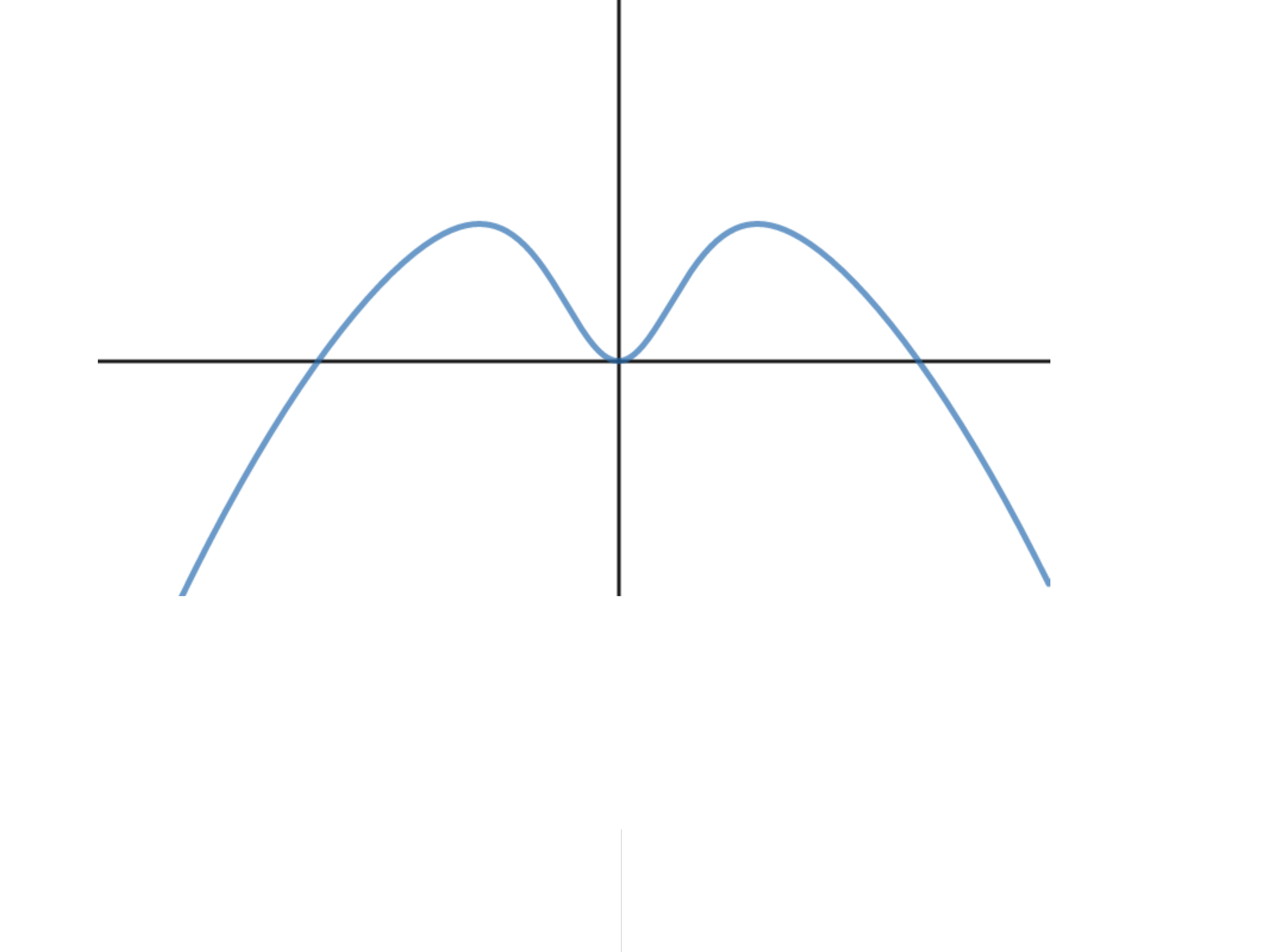}
\end{center}
\vspace{-1.4cm}
\caption{.\label{pitinpotential} The Self Consistent Potential }
\end{figure}

 Consider a density expectation value concentrated near the origin.   Then the equation above is the single body expectation value of the energy of a particle in a potential
 that has the shape of Figure \ref{pitinpotential}.

 It is clear that near the origin, the upside down oscillator potential is negligible.  If we neglect it, then the potential $U$ has bound states concentrated near the origin and the density expectation value in those states can give rise to a self consistent potential concentrated near the origin, if we put a finite fraction of the fermion flavors into those bound states.  
 
 If we now restore the $- \lambda^2$ potential, those states will be meta-stable, with lifetimes of order $1$ in spring constant units.  A meta-stable bound state of $M$ fermions will decay to another meta-stable state of $M - 1$ fermions plus a fermion that propagates out to infinity.   This can be seen most simply by noting that the single particle tunneling probability for a state in one of the bound state wave functions is a number less than one. Of order $N$ fermions must tunnel out in order for the bound state to decay completely.  Alternatively one can calculate the tunneling probability in terms of the functional integral for the Hubbard-Stratonovich field dual to the density operator, which has an action of order $N$ for all semi-classical configurations.
 
 These states exist even at large $\mu$ when the ground state fermi level is far below the maximum of the inverted oscillator potential.  In this case they can be created in scattering only when the energy of the scattering state is $o(N)$ above the ground state.  They will not appear in perturbative scattering at any order of perturbation theory. We emphasize that the perturbative S matrix will be a tensor product of $N$ copies of the $0B$ S matrix, to all orders in perturbation theory.   The exact S matrix will have resonances above a threshold energy of order $N$.  These resonances will be highly degenerate, with the details depending on the spectrum of single particle bound states in the self consistent potential, which depends on $g^2 K(\lambda ,\kappa )$. If there are two or more bound states in the single particle potential, then by putting large numbers of fermions in each, the spectrum of meta-stable levels will be chaotic since they will correspond to large integer linear combinations of irrational numbers. These meta-stable states will decay by the emission of single fermions with a rate that is independent of the entropy of the ensemble of meta-stable states, consistent with the fact that the Hawking temperature of black holes is independent of their mass/entropy in linear dilaton gravity.  The only scales in the model, which determine this rate, are parameters that we've chosen to be $N$ independent in the large $N$ limit.  
 
 Despite these resemblances, this model is not a good model of black hole dynamics.
 The emitted fermions will have a spectrum that obeys Poisson rather than Boltzmann statistics.  Furthermore, the model preserves an exact global $SU(N)$ symmetry, under which the single particle scattering states transform as two copies of the $N$ representation.  The exact S-matrix will obey $SU(N)$ selection rules.
 
 These difficulties are remedied by adding the $J$ term to the Hamiltonian.  It breaks the global symmetry down to a single $U(1)$, and randomizes the dynamics of fermions trapped in the meta-stable states.  The evidence for the latter claim comes from extensive numerical studies of the SYK models\cite{SYKnumerics}, which indicate that those models thermalize rapidly and are fast scramblers.  Indeed, near the origin, one can view this model as an SYK model for a finite number fermion modes trapped near the origin, coupled to a "drain" system that allows fermions to tunnel out to a non-interacting region.
 
 \subsection{Quantum Corrected Entropy Formulae}
 
 Viewing these models as non-relativistic quantum mechanics, it is quite evident that they produce a correct Page curve\cite{page} for an evaporating meta-stable state, as well as a coarse grained thermal spectrum for emission probabilities.  We'll examine this in more detail below. These models also have the right to be called quantum theories of gravity, since
 \begin{itemize}
 
 \item For finite $N$ and large $\mu$ they have a $1/\mu$ expansion for the S-matrix which gives a tensor product of $N$ copies of the S-matrix of the $0B$ string theory, to all orders in the $1/\mu$ expansion.
 
 \item For $\mu = 0$ the fermionic entropy formulae for the ground state predict the form of the dilaton field\cite{tb1+1}, and the finite $\mu$ corrections to this can be expected because of the "stringy" nature of the model.  
 
 \item  The model has meta-stable excitations of high entropy, which are invisible in the $1/\mu$ expansion.  These excitations are thermal states, with a temperature independent of their entropy.  Their entropy and mass are linearly related, because both come from counting fermion flavors.   There is no space-time interpretation of the "black hole interior" .  Recall that in the linear dilaton vacuum there is a mapping between the $\lambda$ coordinate of the non-relativistic fermions and the space-time coordinates. It comes from equating the fermion entropy in a bounded region of $|\lambda|$, calculated in the Thomas-Fermi approximation, to the entropy field $S(r)$.
 We can use that to map an ensemble of our meta-stable states with fixed energy to the black hole solutions of linear dilaton gravity, by setting the value of the dilaton at the black hole horizon to the entropy of the ensemble.  The region inside the horizon is then mapped to the region of $\lambda$ over which the interactions are spread.  The interactions are non-local in $\lambda$ and the "space-time" coordinates are the non-commuting 't Hooft light front variables.  Note that this is consistent with the classical picture of the black hole interior.  The extent of proper time for any time-like trajectory 
between crossing the horizon and hitting the singularity is microscopic and independent of the black hole mass.

\item Although the exact scattering states of the non-relativistic fermion theory are in one to one correspondence with the states of $2N$ Weyl fermion fields in Minkowski space, the model has no ultraviolet divergences at all.

\end{itemize}

The analysis of quantum corrections to classical linear dilaton gravity was done in the seminal paper of Callan {\it et. al.}.  We will repeat that analysis here, differing from them only in a factor of $N$ in front of the classical action, and our notation for the dilaton field.  We emphasize that the string coupling, which "controls" the semi-classical analysis of this model is the inverse square root of the dilaton/entropy field, not large $N$, but that only at large $N$ does the string coupling near the black hole horizon remain small throughout most of the course of evaporation.  The CGHS model is the low energy effective field theory of $N$ copies of 0B string theory, but at $\mu = 0$.  There is no tachyon wall preventing the fermions from entering the region of strong string coupling.  CGHS argued that one could save the semi-classical analysis by taking $N$ large.

From a more fundamental point of view, we see this as evidence that the regime where classical gravitational physics gives a correct description of the underlying model is the regime of large {\it entropy}.
One of the most important lessons of the present paper is that the regime of validity of classical gravity is much larger than that of quantum field theory in curved space-time.  The relationship is quite analogous to that of hydrodynamic equations in condensed matter systems.  The Navier-Stokes and diffusion equations describe long wavelength transport properties of materials in a wide range of states where the phonons of the quantized linearized hydrodynamic equations are not a good description of the microphysics.  

 The classical action of the CGHS model is 
\begin{equation} S_c = N\ \int d^2 x \sqrt{-g} e^{-2\phi}  [R + (\frac{\nabla S}{S})^2 + 4L^{-2}] + \int d^2 x \sqrt{-g}  (\nabla f_a)^2 , \end{equation} where $f_a$ are the bosonized fermi fields and $S = N e^{-2\phi}$.  As pointed out by CGHS, $S$ is the analog for two dimensional gravity of the Bekenstein-Hawking entropy.  In higher dimensions, $S$ is the integral of a $d - 2$ form over a cycle, rather than a pointlike field.  Note that at large $N$, the string coupling $g_S^2 = e^{2\phi}$ becomes strong at singularities in the $\phi$ field, but the entropy at large $N$ is large unless one is very close to the singularity.   The reflection of this in the underlying inverted oscillator model is that most of the entropy is concentrated at large $\lambda$, where we encounter an infinite dimensional space of scattering states.  The entropy is small near the top of the potential, unless $N$ is large.  

From the point of view of higher dimensional Type II string theory, the absence of $S$ in the $f_a$ field action is due to the fact that these are Ramond-Ramond excitations. The one loop effective action contains contributions from the $f_a$ fields and, in principle the gravitational ghosts (since we are going to work in conformal gauge).  While the latter are necessary for a treatment of gravity as a quantum field theory, their contribution to the large $N$ equations is just  an additive shift to $N$ and is subleading for large $N$, the only regime in which we have evidence for black holes\footnote{It is interesting that CGHS came to this conclusion without the insights we've gained since about $1 + 1$ dimensional string theory.  They show that the semi-classical analysis of black hole physics makes sense, if it makes sense at all, only at large $N$.  The reason for this is that the semi-classical expansion parameter $S^{-1/2}$ becomes large near the black hole  horizon, unless $N$ is large.}.   The quantum correction to the effective action includes a bare $S$ independent cosmological constant, which must be set to zero, plus a non-local term
\begin{equation} S_1 = \frac{N}{12} \int \sqrt{- g (x)} \sqrt{- g (y)} R(x) R(y) \nabla^{-2} (x,y) . \end{equation}
There is also a term $ \propto N\int \sqrt{- g} R$, which is an {\it additive shift} in $S$.  Since this term is the same for all states, and has no dynamics, we can drop it.
In conformal gauge the quantum corrected equations of motion take the form

\begin{equation} 0 = N \bigl{[}e^{- 2\phi} (4\partial_{\pm} \phi \partial_{\pm} \rho - 2\partial_{\pm}^2 \phi) + \ T_{\pm\pm} - \frac{1}{12} [\partial_{\pm} \rho \partial_{\pm} \rho - \partial_{\pm}^2 \rho
+ t_{\pm} (\sigma^{\pm} ) ] \bigr{]}. \end{equation}
\begin{equation} 0 = N [e^{-2\phi} (2 \partial_+\partial_- \phi - 4 \partial_+\phi \partial_- \phi - L^{-2} e^{2\rho}) - \frac{1}{12} \partial_+ \partial_- \rho \bigr{]}. \end{equation} 
In these equations
\begin{equation} S = N e^{-2\phi} \ \ \ g_{+-} = - \frac{1}{2} e^{2\rho} \ \ \ T_{\pm\pm} = \frac{1}{2N} \partial_{\pm} f_a \partial_{\pm} f_a. \end{equation}
$f_a$ represent the incoming classical waves of bosonized fermions.
The arbitrary functions $t_{\pm}$ in these equations are a reflection of the non-local nature of the quantum effective action.  They must be fixed by boundary conditions determining the initial and final quantum states of the system.  For details see\cite{cghs}\cite{bddo}\cite{rst1}.

The term $- \frac{N}{12} \partial_+ \partial_- \rho$ represents the stress tensor of the Hawking radiation for black hole solutions.  It vanishes in the linear dilaton vacuum, which is Minkowski space in light front coordinates $\sigma^{\pm}$ and
\begin{equation} \rho = f_a = 0 \ \ \ \ \phi = -(\sigma^+ - \sigma^-)/2L . \end{equation} 
If we drop the quantum corrections, the model is exactly soluble and arbitrary non-zero $f_a$ produces a black hole solution with a singularity, for which the Hawking radiation rate asymptotes to a constant at late time.  Taken literally, this seems to violate energy conservation, mirroring the more recent emphasis on the fact that, since the Hawking particles are entangled with the black hole that emitted them, the entanglement entropy grows linearly at late times.  

The quantum corrected equations of motion, which appear to reflect the correct leading large $N$ behavior of this low energy effective quantum field theory have a conserved stress energy tensor, since they follow from a generally covariant, albeit non-local, effective action.  The authors of\cite{cghs} argued that the solution of this equation would lead to energy conservation.  They speculated that the solution would be non-singular and that all information and energy would be contained in the outgoing radiation, leaving behind the linear dilaton vacuum.

This was shown to be false in\cite{bddo}\cite{rst1}.  The solutions of the large $N$ semiclassical equations is always singular.   The Euler-Lagrange equations of the quantum corrected conformal gauge action have the form of a nonlinear sigma model for $\phi$ and $\rho$ with a target space metric that becomes singular when \begin{equation} e^{-2\phi} = \frac{1}{24} . \end{equation}  At infinity, the classical terms in the effective action dominate and the fields are far from this singularity.   The entropy is large and decreases monotonically as one moves from the conformal boundary of the space-time to smaller causal diamonds\footnote{We order diamonds by the maximal proper time between their tips.}.  In the classical solution, without the Liouville term in the action, the line at which the dilaton crosses the singular line in target space is on the boundary of a causal diamond of finite proper time, of order the Planck time, and the entropy function goes to zero inside that diamond.  This is a conventional space-like singularity cloaked inside a horizon.  

The authors of\cite{bddo}\cite{rst1} showed that a zero entropy singularity was inevitable in the large $N$ CGHS equations.  It is not known (to the present author at least) whether the locus of the singularity in these equations is timelike or spacelike.  Following the work of CGHS a number of authors\cite{cghsmod} investigated modifications of the model, where the semi-classical equations were exactly soluble.   Some of these had no singularity, and in some the singularity became naked; the singularity locus turned time-like on some Cauchy slice.   Generic boundary conditions on the time-like portion lead to models with energy unbounded from below.  This is also the fate of singularity free solutions.   Black holes continue to evaporate forever.  These models are inconsistent with black hole thermodynamics, or any interpretation in terms of a consistent quantum theory with a Hamiltonian bounded from below.  Generalizing to models that are not exactly soluble another possibility is that the singularity remains spacelike in the quantum corrected equations, but that the Hawking flux goes to zero.  In any such model, the entropy on the quantum corrected horizon will be order $N$, unless the value of $e^{-2\phi}$ is of order $1/N$.  Since $N$ does not appear in the equations for $\phi$ and $\rho$, this indicates a singularity on the horizon.  The linear dilaton vacuum indeed has such a null singularity, but there is no other solution with this property.   Thus, a solution of this type would have a zero energy remnant, with entropy of order $N$.

The authors of\cite{rst3} propose a boundary condition on the time-like portion of the zero entropy singularity in their soluble semi-classical model, that ends the singularity after a very short proper time and joins it to the null singularity of the linear dilaton vacuum solution.  This is sensible if all but an order $1$ amount of the original $o(N)$ black hole entropy is emitted to infinity to the past of the null line joining the point where the singularity curve turns timelike, to future null infinity.  RST call this the {\it thunderpop} boundary condition, following a suggestion of Hawking that a macroscopic amount of energy could be released in a thunderbolt if the black hole singularity became naked. The three possibilities of a remnant, a thunderbolt and a thunderpop are illustrated in Figure 2.
\begin{figure}[h!]
\begin{center}
 \includegraphics[width=12cm]{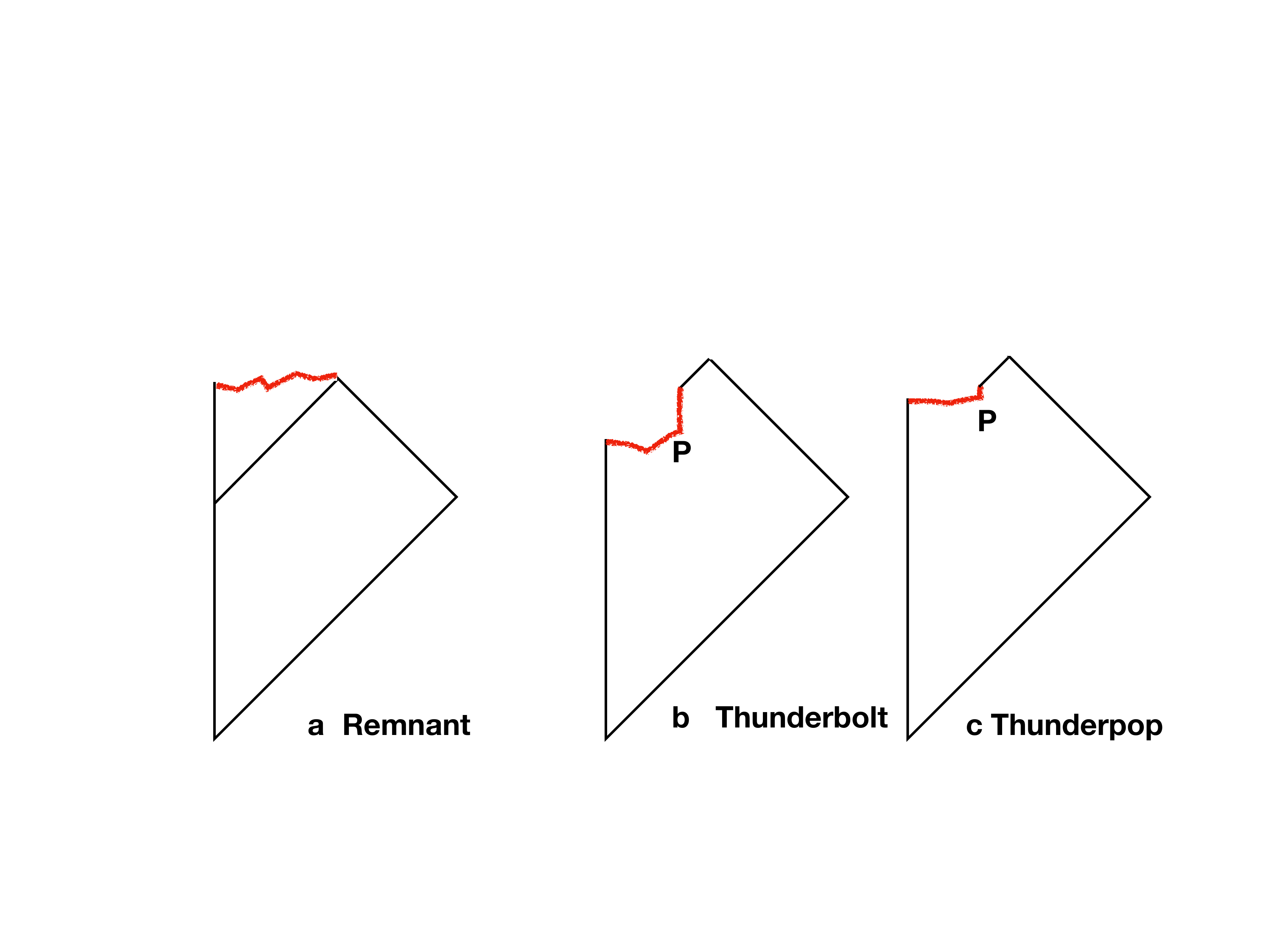}
\end{center}
\vspace{-1.4cm}
\caption{.\label{3poss} Three Scenarios for Linear Dilaton Black Hole Evaporation }
\end{figure} 

The QES proposal attempts to improve upon the Page or entropy flow curve that is suggested by the lowest order semi-classical estimate of Hawking radiation.  Page's argument focused on the entanglement entropy of outgoing radiation with the black hole.  Obviously, this grows with time from the moment the black hole forms and begins to radiate. In the naive semi-classical calculation the black hole radiates forever and the growth of entanglement continuous indefinitely.  However, for a typical pure state of a large bipartite system the entanglement entropy of the larger Hilbert space becomes negligible compared to its entropy.

In\cite{lindilisland} the authors propose a version of the quantum extremal surface/island proposal appropriate to linear dilaton models.  The second of these two references works explicitly with the model of\cite{rst23}.  We reproduce here the Penrose diagram for their island proposal 

\begin{figure}[h!]
\begin{center}
 \includegraphics[width=12cm]{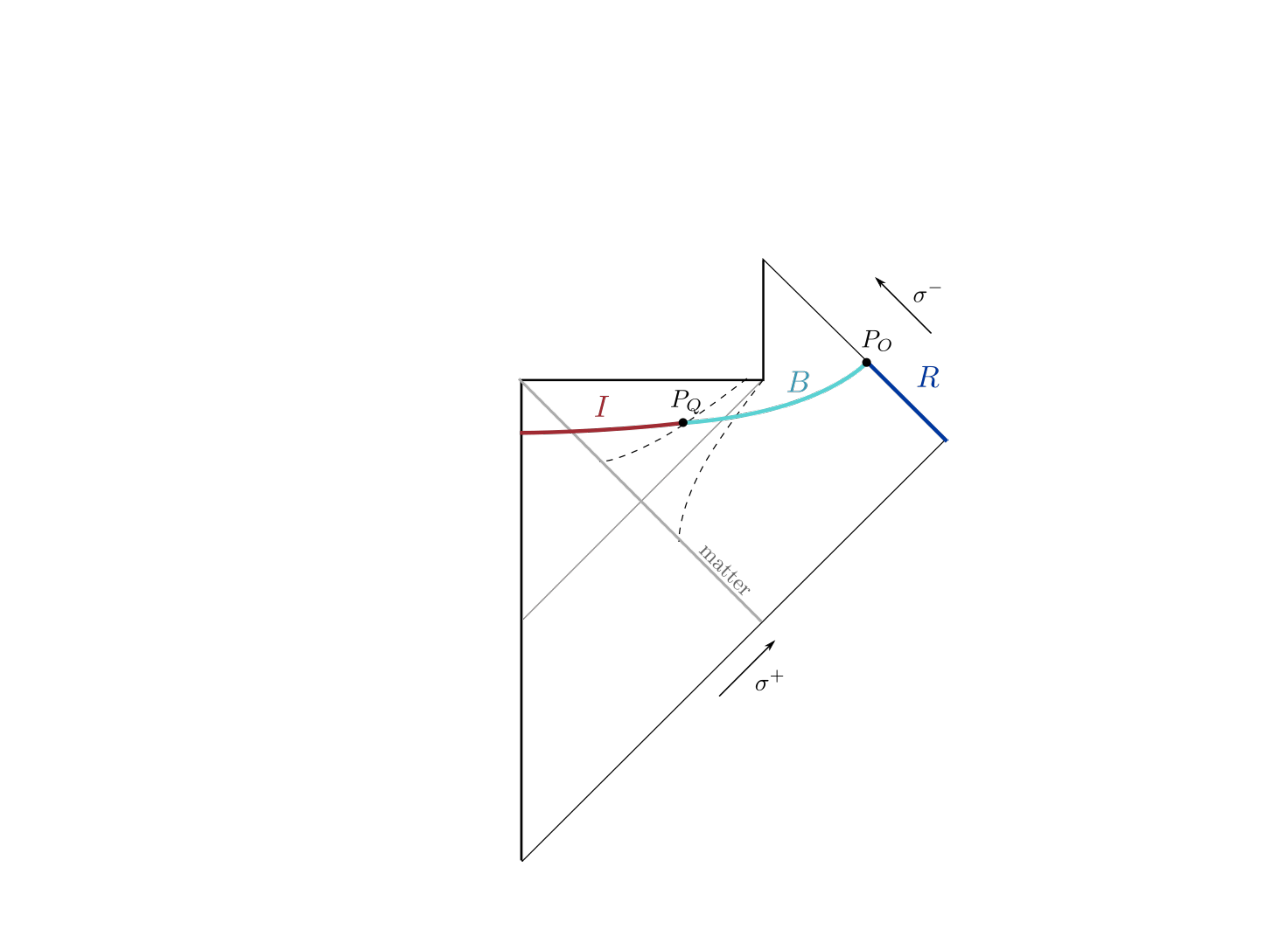}
\end{center}
\vspace{-1.4cm}
\caption{.\label{lindilisland} The Island Proposal for CGHS/RST Models }
\end{figure} 

The point $P_Q$ is inside the black hole horizon. Draw a spacelike slice through $P_Q$ which asymptotes to a point on future null infinity.
 The generalized entropy is the sum of the classical entropy function $S(P_Q)$ evaluated at this point, the entanglement entropy of the radiation that has been emitted to the past of that space-like slice and the entanglement entropy of quantum fields just outside $P_Q$, with the interior.  Equivalently one is evaluating the entanglement entropy of quantum fields in the space-like interval between $P_Q$ and null infinity.  Note that $S(P_Q)$ is evaluated on the solution of the large $N$ semi-classical equations, not the classical linear dilaton black hole.
 Because of the latter fact this prescription is slightly different than the one invoked in the AdS/CFT context.  One then extremizes the generalized entropy with respect to $P_Q$, and the extremum is the QES  for this model. More precisely, there are several extrema and the QES is the one of minimal generalized entropy.  One then evaluates the generalized entropy as a function of the coordinate on future null infinity.  For comparison with calculations in the UV complete model recall that there is a map between asymptotic null coordinates and $\lambda$, such that sending the null coordinate to the future corresponds to increasing $|\lambda|$.

The evolution of the quantum extremal surface is such that in the model of\cite{rst23} it approaches the point where the zero entropy curve turns timelike, as one proceeds toward future infinity on the null boundary.  For large mass black holes, the generalized entropy approaches zero to leading order in the mass, at precisely this point.  It is unclear how one would define the island prescription to the future of the null line emanating from the point where the singularity turns timelike.  

Although the authors refer to the "thunderpop" boundary condition of\cite{rst3}, their quantum extremal surface and island computation of the entanglement entropy of the outgoing radiation takes place on a Cauchy slice prior to the appearance of the naked singularity.   Thus, their computation does not distinguish between models that actually have a Page curve, and models where one expects to violate at least the thermodynamic interpretation of gravitational fields and probably the rules of quantum mechanics. The appendix to\cite{hss} contains an attempt to prove the island formula by the calculation of "replica wormholes"\cite{replicawormholes} in the model of\cite{rst3}.  It is possible that this calculation will fail if one does not impose the "thunderpop" boundary condition.

To summarize: a rather general entropic analysis of variations on the CGHS model, using only the principle that the entropy function in a nested sequence of causal diamonds should be a monotonic function of their proper times, suggests only three possible outcomes for these models.  If the zero entropy singularity is space-like, then the large $N$ equations will contain a large entropy remnant with microscopic energy.  If it turns time-like at some point $P$, then the black hole will radiate down to negative energy.  If all but an order $1/N$ fraction of the positive energy of the initial state is radiated to future null infinity prior to the null line joining $P$ to the conformal boundary, then we can exploit the necessity to impose boundary conditions on the naked singularity, to join it continuously to the null singularity of the linear dilaton vacuum.  This provides the basis for proposing a Page curve for the entanglement entropy of Hawking radiation in these models, and the island formula of\cite{lindilisland} provides a calculation of it.  This calculation is independent of the choice of boundary conditions on the naked singularity and so cannot by itself be considered a semi-classical understanding of the Page curve, since it is applicable to models in which the Page curve is not the correct semi-classical answer.  It is possible that the replica wormhole derivation of the island formula will depend in a crucial manner on the imposition of the thunderpop boundary condition in the Lorentzian model.  This would certainly be the most satisfactory outcome and would validate the entire circle of ideas surrounding quantum extremal surfaces, islands and replica wormholes.

To conclude this subsection, we will propose some optimistic conjectures for the general class of dilaton-gravity models given by the Lagrangians
\begin{equation} {\cal L} = \sqrt{-g} [S R - F(S) (\nabla S)^2 - U(S)] . \end{equation}  Let us insist that the equations of motion have a solution with a global time-like Killing vector that asymptotes on the boundaries of large causal diamonds to the linear dilaton solution of the CGHS model.  The range of the field $S$ is $[0,\infty]$.
This requires that at large $S$, $F$ and $U$ become linear.
We also add $N \gg 1$ bosonized fermion fields,$f_a$ minimally coupled to the metric, but not to $S$.   Our first conjecture is that for some class of functions $F,G$ there is a positive energy theorem for the large $N$ semi-classical equations for these models.  Note that, in order to write $N$ independent equations we take $S = N e^{-2\phi} $ so that only the leading order behavior of the functions $F,G$ survive in the large $N$ limit.  The unique surviving class of Lagrangians is that of\cite{rst3} with an arbitrary coefficient of the ${\rm ln}\phi $ term.   The second conjecture is that for those models the solution of the semi-classical equations with incoming $f$ waves always has a space-like zero entropy surface, cloaked by a horizon of finite entropy, which will be of order N.  The replica wormhole derivation of the island formula should lead to the conclusion that the outgoing radiation is entangled with the remnant.  It may be that this class of solutions is empty.

Models without a large $N$ positive energy theorem will divide into two classes depending on whether there is a zero entropy singularity or not.
Those without singularity are conjectured to be unstable.  Models in which the zero entropy singularity turns timelike at a point $P$ can be saved if we impose the thunderpop boundary conditions at a point before the energy radiated to infinity exceeds the energy in the black hole.  Note that the question of whether singularities exist depends on corrections to the functions $F$ and $U$ at small $S$, where the singularities occur.  One can also imagine models in which there is a macroscopic amount of energy radiated prior to a null ray stretching from $P$ to the conformal boundary.  This would be a Hawking thunderbolt.  Such models would be bizarre because they seem to localize a large amount of energy in a space-time region of very low entropy. The proposal of\cite{lindilisland} would not distinguish between such models and the sensible thunderpops, unless the replica wormhole computation fails.
The authors of\cite{hss} have verified that the model they have studied has a thunderpop rather than a thunderbolt.  However, the location of the point $P$ depends on the behavior of $F$ and $U$ at small $S$, so we need an argument, presumably based on replica wormholes, that thunderbolt models have a different island rule, or one that shows that all but a microscopic amount of energy is released prior to $P$, for general choice of $F$ and $U$.

\subsection{The Page Curve in UV Complete Models}

 It is worthwhile going through the detailed steps of time evolution corresponding to black hole formation and evaporation in the UV complete models. We will work in the language of the non-relativistic fermions.
We begin from the obvious remark that our system is a unitary quantum model, so the "black hole state" that we produce by starting with an asymptotic  state containing left moving wave packets of $M\gg 1$ flavors of fermion, is pure.  What we mean by the black hole entropy is the following. Once the fermions reach the top of the potential there are of order $2^{2M}$ meta-stable states of the model with $J = 0$, in which $\sim M$  fermions are bound to the dip in the potential.  It will be illuminating to think about the parameter regime $1 \ll M \ll N$.
The $J$ couplings make rapid transitions between these states, scrambling the information in a time of order ${\rm ln M}$ in units of the oscillator spring constant.
Scrambling means {\it classical thermalization}.  The modern understanding of that term\cite{deutschsrednickiberry} is that there is a large class of "simple" operators, which one may call ETH operators, whose time dependent correlators evaluated in the Heisenberg picture state defined by the incoming boundary conditions, is the same, up to corrections down by inverse powers of the entropy, as they would be in a random state taken from the thermal ensemble with the same expectation value of the energy as the Heisenberg state\footnote{The Heisenberg state consists of wave packets and is not an eigenstate of the energy.}.  It is {\it not true} and will {\it never be true for this system}, that {\it arbitrary} operators take close to the same value in the Heisenberg state as they do in the ensemble of black hole states.  The black hole decays long before its quantum state becomes truly random. We define the black hole entropy to be the von Neumann entropy of the ensemble of states for which this subset of ETH operators takes on the same values.
Very general arguments show that it is a thermal ensemble, with temperature determined by the expectation value of energy in the initial state.  It is also clear in this model that for times of order the scrambling time, the multibody wave functions of most of the fermions in all of the meta-stable states fall off exponentially beyond some value $|\lambda| > \lambda_* \sim 1$ in spring constant units.

The dynamics of our model at large $N$ is simple enough to  analyze what happens to a generic state in the meta-stable ensemble, because the Hartree approximation is valid at large $N$, as a consequence of large $N$ factorization.  Consider a generic single fermion meta-stable state $\psi_a (\lambda)$ .  It will tunnel out to the continuum with a probability $< 1$ but order $1$.  The general state in the Hilbert space of fermions with wave functions restricted to be within range of the potential, is a superposition of products of these single fermion wave functions.  We claim that its tunneling probability is well approximated by the probability of being in a particular product state times the decay probability of that product state.  What we mean by this is that, if we only measure simple operators, like the number of fermions of each type that come out with a given single particle energy, then the relative phases between different product states are irrelevant.   In this approximation it is clear that the answer is going to be thermal.  

On the other hand, it is also clear that, with enough computational power we could calculate the exact S-matrix for this model and verify that it is unitary.  If we had that in hand, we would never attribute entropy to the black hole created in a given experiment {\it if we had complete knowledge of the initial physical state}.  This means: we know the projection of that particular black hole state onto the entire subspace of final states of all possible black hole states in the ensemble.  This subspace has dimension that is a power greater than $1$ of $e^S$\cite{zurek}\footnote{There is a question of what we mean by $S$ in this formula.  The most precise definition would be to specify an energy interval around the average energy of the initial state and define $S$ to be the microcanonical entropy of that subspace.}
, unless we are able to do enough measurements on the final state to exactly determine what it is, and not simply its coarse grained properties.  This requires knowledge of the S matrix.   If, for example, we knew only that the correct model of the world was one of the many we've exhibited in this paper, which have black hole excitations, we would have to either do the computation for every single model, find out what final states are produced from the subspace of scattering states that produce black holes (which will be different for each model) and compare all of these computations to experimental predictions.  Quantum mechanics is probabilistic, which means that experimental checks require multiple measurements of at least two non-commuting operators on a space of dimension $e^S$  with non-degenerate spectra, with care taken that the initial state in each experiment is exactly identical. 

In short, while the current set of models are all certain to have unitary S matrices for black hole formation and evaporation, the statement that a particular black hole state has vanishing entropy is free of operational content.   In any conceivable experiment, the initial and final states will be mixed.  That is to say, we will have old fashioned pre-quantum mechanical uncertainty in what both the initial state is that creates the black hole, and in the final state of Hawking radiation from the evaporated black hole.  We will be computing transition amplitudes between two density matrices whose entropy is at least as large as the entropy of the normalized projector on the subspace of states in 
a microcanonical energy band around the black hole mass.  
 
In the present model, the nearest one can come to a fundamental computation of the sort contemplated in\cite{qes}, is the following:   Choose a point $\lambda_{0}$ outside the range of interaction kernel $K(\lambda, \kappa)$.   Prepare an initial pure state, which will form a black hole, and a time $t_0$ after black hole formation but long before its Page time.  At that time, compute the sum of the entanglement entropy of fermions confined to the interval $\lambda > \lambda_0$ and the black hole entropy function "at the point $r_0$ corresponding to $\lambda_0$", in the Schrodinger picture state at times $t > t_0$.  
The phrase in scare quotes is problematic.  As shown by\cite{dasetal}, the entropy formula gets corrections {\it at weak string coupling}, which muddle the identification of $\lambda$ and $r$.  Indeed, this should be expected from the results of\cite{tedetal}: geometric entropy formulae get corrections from higher derivative terms in the action, and in weakly coupled string theory these set in at length scales much longer than the Planck scale.  As argued in\cite{tb1+1}, the Planck scale in these models is the scale set by the upside down oscillator spring constant.  It is the scale that remains in the model when we make the string coupling of order $1$.   This means that in weakly coupled string theory, the point $\lambda_c$ where we are allowed to use the linear dilaton Lagrangian in our calculation of the relation between $r$ and $\lambda$ satisfies $\lambda_c \gg \lambda_0 $, for any choice of $\lambda_0$ close to the support of the meta-stable state wave function.  

This difficulty is compounded with a second one, which also mirrors a problem in the standard calculation of entanglement entropy in effective field theory, namely that it is always divergent.  There are no UV divergences in our model\footnote{The "UV" infinities here are all associated with the negative energy Dirac sea and can be cut off as in\cite{moore} by putting boundary conditions at very large $\lambda$.  This cutoff does not affect the discussion at all. Moore's calculation can be viewed as the first example of the UV/IR connection familiar from AdS/CFT.}, but there is an entanglement entropy of fermions with $|\lambda| > \lambda_0$ in the ground state of the system.   It is of order $N$ and is much larger than the black hole entropy for a black hole produced by sending in only $M$ fermion flavors with $1 \ll M \ll N$. This ground state entropy can be computed exactly only for $g^2 = J_{abcd} = 0$, and it depends on the string coupling (the fermi level $\mu$) .    In the effective field theory calculation it should be thought of as the $UV$ divergent part of the calculation.  Our UV complete theory shows that it is non-universal and depends on multiple parameters. One would need the complete solution of the model with finite $g^2$ and $J_{abcd}$, at least at large $N$, to calculate it.   Also, unlike the UV divergent effective field theory 
entanglement entropy, it depends on $\lambda_0$.  The authors of\cite{hartnoll} showed that the entanglement entropy of fermions in an interval on the $\lambda$ axis, exactly matched the effective field theory computation of entanglement, but only for very large $\lambda$ where stringy corrections to the relation between effective field theory and the actual model always become negligible.  Their computation is easy to understand in terms of the transformation of\cite{akk}.   

At strong string coupling, when the fermi surface is at the top of the potential, then the map between $\lambda$ and $r$ in the $g^2 = J_{abcd} = 0$ model does follow the classical Lagrangian.  This will be modified in models with black holes, but only for $\lambda_0$ close to the top of the potential, where $K(\lambda, \kappa)$ is supported.  Thus, the ground state subtraction and the relation between $\lambda_0$ and $r_0$ are unambiguous, as long as $\lambda_0$ is outside the range of the interaction, and beyond the range where stringy effects are important.  The answer depends on the string coupling and also on the interactions in the interior of the black hole.
It approaches the effective field theory computation only at very large $\lambda_0$.

We can now follow the evaporation of the black hole for $t > t_0$.
If the original black hole state was prepared with $M$ fermions, then, in the Hartree approximation, the wave function is a product of $m_1$ fermions of different flavor in the first quasi-bound state, $m_2$ in the second, {\it etc.} , with antisymmetrization imposed in a trivial manner by summing over which fermions are in which quasi-bound state.  The $J$ couplings will, of course, produce a more complicated linear combination of these Slater-Determinants as approximate eigenstates of the Hamiltonian.  The decay amplitudes of the linear combinations are normalized linear combinations of the decay amplitudes of Slater states, and this makes it clear that the probability for emitting more that a few fermions per unit time that can propagate into the region $|\lambda| > \lambda_0$ is highly suppressed.  This implies that the rate of energy loss of the black hole is linear in time, independent of its mass, in agreement with Hawking's calculation and black hole thermodynamics.      It is also obvious that the subtracted entanglement entropy between the emitted fermions and those that have not yet been emitted grows linearly with time, reaches a maximum, and then saturates at some finite value.  That value cannot be calculated without knowing the microscopic details of the interaction, in order to properly define the large subtraction, unless one takes $\lambda_0$ extremely large.  At large $\lambda_0$, the modifications of the fermion wave functions due to the $g^2,J$ interactions give only exponentially small corrections to the vacuum entanglement.  
For black holes of entropy $1 \ll S \ll N$, the calculation of the vacuum subtraction must be done with an accuracy finer than $S/N$\footnote{  This suggests that methods based on averaging over the couplings $J_{abcd}$ may not give the right result. }.  If we want to use the effective field theory formula for the entanglement entropy this means that we need to choose $\lambda_0 \gg N/S$.   A scrambling time after the meta-stable state has formed there is virtually no entanglement between states of fermions in the meta-stable wavefunctions, and those with wave functions concentrated in the region $|\lambda| > \lambda_0$ .   This is a consequence of the exponential falloff of the initial meta-stable wave functions away from $\lambda = 0$.  

Decay proceeds by single fermion tunneling out of the potential well at the origin.  After some time, the wave function of this first Hawking fermion will be concentrated in the region $|\lambda | > \lambda_0$.  Since this fermion was entangled with the other $o(M)$ fermions in the meta-stable state, this will increase the entanglement entropy of the asymptotic region.   Mathematically what we are saying is that the multibody wave function of the meta-stable state is a superposition of products of single particle wave functions.  The single fermion tunneling out could be any of the fermions in the bound state.  Eventually the wave function looks like $$\sum_n \Psi_n (\lambda_i ) \psi_n (\lambda) $$, where the $\psi_n $ are concentrated at large $\lambda$ and the multibody wave functions
$\Psi_n$ at small $\lambda$ .  The label $n$ runs over both the fermion flavor label and some basis of localized orthonormal functions centered at large $\lambda$.  As time goes on, more fermions tunnel through the barrier and the wave function has the form $$\sum_n  \Psi_ n (\lambda_< ) \tilde{\Psi}_n (\lambda_> )$$ where the index now runs over a class of multi-body wave functions concentrated at coordinates larger than and smaller than $\lambda_0$.  It is clear that eventually the entanglement entropy of the large $\lambda$ fermions
with those at small $\lambda$ will go to zero.  This is just Page's theorem, applied to a finite dimensional subspace of states in a completely finite non-relativistic system of particles.  The calculations of\cite{hartnoll} and the results of\cite{akk} show that if we take $\lambda_0$ sufficiently large, then the entanglement entropy of the large $\lambda$ fermions will be that calculated in the effective CGHS field theory, up to terms exponentially small in $\lambda_0$.  In order to completely validate the island rule calculation one would have to show that the large correction to the vacuum entanglement, which cannot be calculated in effective field theory, did not affect the minimization of the quantum RT formula with respect to the value of $r_0$ that corresponds to $\lambda_0$.  This is plausible for large $\lambda_0$ but is not completely obvious in the weak string coupling regime, where the corrections are power law, rather than exponential.

{\it In contrast to the entanglement between fermions at large and small $\lambda$, the entanglement, after the Page time, between the large $\lambda$ fermions, can have nothing to do with low energy effective field theory.  Knowing this is tantamount to knowing the black hole S matrix, which, in our models depends on a large number of couplings that are invisible not only in the low energy field theory, but to all orders in string perturbation theory.}

Another thing that cannot be correct in this model is a picture in which the calculation of the Page curve is telling us something about "an island in the black hole interior".  In the classical black hole geometry, the proper time for any geodesic to hit the singularity is order Planck time, the scale set by the inverted oscillator constant.  General folklore would tell us to abandon the general relativistic space-time picture in the black hole interior.  Our explicit UV complete models show us that this is the case.  The black hole interior is identified with the region in $\lambda$ space where the $g^2$ and $J_{abcd}$ couplings have strengths of order $1$.  There is no inkling of the nature of these couplings in the linear dilaton Lagrangian, or indeed at any order in string perturbation theory.  In order to access these couplings in a regime that has a space-time interpretation in the models, all we can do is calculate the S matrix.
This makes the part of the island rule that involves calculating field theory entropy near the point $P_Q$ in the interior of the black hole somewhat suspect.  

Yet another indication that low energy methods are missing a lot of important information about the physics of black holes is the fate of the $SU(N)$ symmetry of the CGHS model, which is shared by the model of\cite{rst23} and is valid to all orders of string perturbation theory for our models.  Suppose we choose an in state consisting of $1 \ll K \ll N$ fermions, and transforming {\it e.g.} in the rank $K$ anti-symmetric tensor of $SU(N)$, and a smaller number $k \sim 1$ transforming in some representation in the $k$-fold tensor product of the $[N]$ representation.   Then the out state will have zero fermion number for each of the $N - K - 1$ fermion flavors not included in the initial state, and will transform in the same represention of $SU(N)$.  In our models, the fate of this symmetry depends on whether or not the $J$ couplings are present.  We argued that those couplings were necessary to insure the scrambling and thermalization properties of black holes.  If $J_{abcd} \neq 0$ the $SU(N)$ selection rules on S-matrix elements will not be valid. Since we can take $K$ very large, this is a macroscopic effect, and shows that the $J = 0$ model does not have a thermal spectrum.
We see another indication that neither low energy effective field theory calculations nor string perturbation theory can properly account for black hole physics.

The fact that methods based on QES, and the island formula that devolves from it, do appear to reproduce the correct form of the Page curve, might be a consequence of the fact that these ideas, despite appearances, do not represent simple one loop calculations in effective gravitational field theory.  The papers\cite{replicawormholes} show that the derivation of the island prescription from the bulk gravitational path integral dual to the SYK model actually involves a non-perturbative "replica wormhole" contribution.   While it is quite clear that this calculation cannot capture either stringy effects at weak string coupling in linear dilaton models, or the details of particular models in the large ensemble of SYK couplings, it may be capturing certain coarse grained features of the non-perturbative physics.  

\subsection{ER = EPR?}

Eternal black hole solutions and de Sitter space all have Kruskal extensions, with two infinite proper time causal diamonds connected by an Einstein-Rosen (ER) bridge.  The two diamonds are connected by a time reversal transformation, and all of these are thermal systems.  Werner Israel suggested in 1976\cite{israel} that the Kruskal geometry of the Schwarzschild space-time represented the thermofield double of a quantum system.  This suggestion was generalized to AdS black holes by Maldacena\cite{maldathermo}, and to dS space by the authors of\cite{sussetal}.  In this picture, the physical system is restricted to a single causal diamond, the time reversed diamond is the thermofield double copy that purifies the thermal state, and the ER bridge between the two diamonds represents the entanglement between the two systems.  Maldacena and Susskind\cite{erepr} suggested a radical generalization of this idea, called {\it ER = EPR} in which any kind of quantum entanglement was said to be represented by ER bridges that might be microscopically small.  When the entanglement between two systems is caused by actual coupling in the Hamiltonian, the "wormhole becomes traversable"\cite{traversable}.  While the ER = EPR paradigm is useful in its original context, the question raised by this proposal is really one of utility, rather than principle.   It is similar to the question of whether gravity in AdS space is a useful description of anything in maximally supersymmetric Yang Mills theory with small gauge group.

It seems evident to the present author that if the quantum model dual to the space-time had been presented as a boundary theory, which is to say only as an S-matrix theory, that applying the ER=EPR paradigm, one would have interpreted the coupled $N$ fermion model as living on $N$ copies of the linear dilaton space-time, coupled by a set of $N(N - 1)/2$ pairwise traversable wormholes.  It is only because of the map between the space-time picture and the $\lambda$ coordinate that we avoid this mistake. It is always possible to preserve the ER=EPR slogan by insisting that wormholes are microscopic and traversable. What is definitely true in the present model is that this picture of the space-time is not a useful one for understanding the physics.  ER=EPR is a completely valid and useful idea for understanding the quantum physics of weakly curved space-times with a pair of disjoint causal diamonds with opposite time orientation and infinite proper time.  It can be extended to the case of traversable wormholes by reversing the time orientation of one of the diamonds, and coupling the thermofield double systems together.  The present models should serve as a warning for attempts to extend the validity of the hypothesis beyond its original domain.  

\subsection{Firewalls}

As already pointed out in\cite{tb1+1}, the models of this paper cannot resolve the firewall\cite{fw} controversy.  The region behind the black hole horizon is of Planck scale and the large smooth region behind the horizon, which is claimed to be in conflict with the monogamy of entanglement, does not exist.   These models do however contain elements that are consistent with the resolution of the firewall paradox posed in\cite{mirage}, and falsify some of the assumptions on which the firewall conundrum is based.  In\cite{tb1+1} and\cite{thermalization}, the author pointed out that adding a small mass to a linear dilaton black hole, increases the entropy by a small amount and that this is entirely consistent with the idea that the large smooth region behind the horizon of higher dimensional negative specific heat black holes is a hydrodynamic image of the fact that dropping a small low entropy mass into a large black hole leads to a huge increase in the entropy of the final equilibrium state.  The principles of quantum mechanics then tell us that prior to equilibration, many degrees of freedom must be frozen, even though the Hilbert space is much larger.  It is a small step from there to postulate that the frozen degrees of freedom are precisely those that mediate interactions between the small system and the black hole.   If the time scale for the Hamiltonian that equilibrates the system is the horizon crossing time, then the small system will not feel the presence of the black hole for a time of that order.  A generic class of models with this sort of dynamics are models where functions of the fundamental variables are organized into matrices and the Hamiltonian is a single trace of a polynomial.  Then constraints that set off diagonal matrix elements to zero, decouple diagonal blocks.  We'll refer the reader to\cite{mirage}\cite{thermalization}\cite{newton} for details.  If this is the mechanism that explains firewalls, then the entropy formula for linear dilaton black holes tells us not to expect a large smooth interior.

The second piece of firewall folklore that is falsified by these models is the key idea that there is a unique low energy state "in the vicinity of the horizon".  In the previous section we discussed the UV complete calculation of entanglement entropy of the fermion fields with degrees of freedom in the interior of the black hole.   That calculation shows that relativistic fermion fields, which for large enough $|\lambda|$ are linear functionals of the non-relativistic fermions, are entangled with degrees of freedom whose Hamiltonian is a fast scrambler, and which do not have a local space-time interpretation.

It must be admitted that the implication of this calculation for higher dimensional black holes is ambiguous. They could be interpreted as evidence for the existence for a firewall, which forms immediately after horizon crossing.  However, in\cite{tb1+1} we presented a model of the near horizon region of higher dimensional, negative specific heat, black holes with no firewall.  That model does not have an asymptotic region.  It uses only a finite portion of the $\lambda$ axis and pretends to describe the "near horizon" part of the S-matrix.  Similar ideas have been explored in\cite{panos}.  We will illustrate the model only in space-time dimension $d = 4$ .  Higher dimension analogs can be constructed using the formalism outlined in\cite{tbjk}\cite{newton}.  The fundamental variables are fermions, as in the linear dilaton model, and are labeled $\psi_{ab}$, an imaginary anti-symmetric $N\times N$ matrix, whose off diagonal blocks transform in the $[2,1] \oplus [1,2]$ representation of the $SU(2)$ spinor bundle over the two sphere, with a cutoff of the angular momentum
at $N - 1/2$.  The matrix elements of $\psi$ are independent fermionic oscillators, with the superdiagonal elements being Hermitian conjugates of the subdiagonal elements.
The Hamiltonian is
\begin{equation}  \frac{1}{N {\bf L}^2} {\rm Tr}\ P(\psi^2 /N)  + H_m + G({\bf L^2}), \end{equation} where $P$ is a finite polynomial.  ${\bf L^2}$ is the total squared angular momentum of the system.  $G$ is a function that grows linearly at infinity and is $ < 1/N$ for ${\bf L}^2 \ll N^2$.   Its role is explained in Appendix B.
"Scattering states" are defined as in the linear dilaton model.  $H_m$ is chosen to have energy scales of order $1/m$, and is of the form $PH_mP$ where $P$ is the projector on the constrained subspace defined below. 
We thus work in a Hilbert space with maximal entropy ${\rm ln}(2) N(N - 1)/2$ where $N \propto (M + m + \epsilon)$.  $M$ and $m$ are the masses of the black hole and infalling system in Planck units and  $\epsilon$ is inserted in order that the maximal entropy is larger than that of the final equilibrated black hole.  In a loose way we can think of this as the Hilbert space appropriate to a detector on an accelerated trajectory that gets close to the black hole.  

The initial constrained state is taken to be one in which the off-diagonal matrix elements of $\psi^2$ between a block of size $m$ and a block of size $M$, vanish.  More precisely, the superdiagonal components of $\psi$ are annihilation operators, which vanish acting to the right and the subdiagonal components are creation operators, vanishing when they act to the left.  As a consequence, in the initial state there is no interaction between the fermions in the two blocks.  The normalization of the Hamiltonian is chosen so that the natural time scale for turning on the off diagonal fermions is $\sim N$, while the Hamiltonian is almost certainly a fast scrambler\cite{sekinosusskind}.  Intuitively this is because the trace terms are invariant under the fuzzy remnant of the group of area preserving maps of the sphere.  The square of the angular momentum in the denominator of the Hamiltonian, which breaks area preserving invariance, is inserted in order to produce contributions to the scattering matrix matching those of\cite{thooft}.  Despite the association of 't Hooft's scattering matrix with quantum chaos\cite{sspolch}, this is not by itself sufficient to produce fast homogenization of information on the fuzzy sphere.  The matrix form of our Hamiltonian is what assures that.

Since the $(L^2)^{-1}$ factor in the Hamiltonian breaks area preserving map invariance, we must specify {\it which} matrix elements are being set to zero.  Since $N$ is large, we can construct good approximations to sections of the spinor bundle that are proportional to characteristic functions of subsets of the sphere.  We choose the degrees of freedom in the $m$ block to be small outside of a spherical cap of radius $\sim m$ on a round sphere of radius $\sim N$.   The vanishing elements are chosen to 
be spinor sections small outside of an annulus of area $Nm$ surrounding that spherical cap.  Spinor bilinears are thought of as differential forms on the sphere and the trace of polynomials in these bilinears is the integral of products of forms, so we get a geometric (really measure theoretic) interpretation of the lack of interaction between the $m$ and $M$ blocks, in the initial state.

Fast scrambling implies that in a time of order $N {\rm ln}\ N$ the initial state constraints have been erased and the state resembles a thermal state for simple operators.  It is easy to see\cite{unpub} (see Appendix B) that in such a state the average squared angular momentum vanishes and the uncertainty in $L^2$ can be made small by an appropriate choice of $G$.   Thus, the initial localized perturbation is homogenized in a scrambling time.  On the other hand, during the long equilibration time, the $m$ subsystem evolves much more rapidly under the  influence of $H_m$.  It behaves initially as if the larger subsystem were absent, only gradually becoming equilibrated with it by losing degrees of freedom a few at a time.
It is this evolution that\cite{mirage} claimed mirrored the slowly varying part of the black hole interior, in which causal diamonds along the trajectory of the infalling system slowly shrink to zero.  

Hawking radiation occurs in this model when the thermal equilibrium state with entropy $\sim N^2$ fluctuates into a state where the off diagonal matrix elements between a block of size $E$ and the rest of the matrix $\psi^2$ vanish.  The thermal probability that the quantum state will have an order $1$ projection on a state satisfying $\sim NE$ constraints is $e^{- c N E}$, where $c$ is a model dependent constant.  Since most of the accessible states of the system have very small angular momentum, and the fixed $L^2$ Hamiltonian is invariant under fuzzy volume preserving maps, we can think of the typical state of this type as a spherically symmetric superposition of states in which the variables vanish in an annulus of area $\sim NE$ surrounding a spherical cap of area $E^2$.  

We can now address the question of which degrees of freedom are entangled with the outgoing Hawking radiation.  As a consequence of scrambling this is a hard question to answer in detail after the scrambling time.  The coarse grained answer is given by Page's theorem\cite{page}.  The Hawking radiation is a random subsystem, of very small size, of the Hilbert space of maximal entropy $(M + m)^2$ and is entangled with only a very small number of degrees of freedom.  The probability that some of those degrees of freedom come from the subsystem $m$ is of order $e^{ - (M^2 - m^2)}$.  Thus, despite the fact that the $m$ subsystem experiences "no drama at the horizon", there is no field theoretic description of the entanglement of the radiation.  The claim  that the states "behind the horizon" are well described by field theory, is completely wrong in this model.   It is only the states of the $m$ subsystem for which that can be said, and only for a time of order $N$.  Once the scrambling time has passed, there is no black hole interior.  If a new small subsystem is dropped onto the equilibrated black hole "the interior is recreated" for a brief period, but it is experienced only by a small subset of the behind the horizon degrees of freedom.  An external detector sees this equilibration process as the decay of perturbations on the horizon.
These considerations resolve the firewall paradox, as originally suggested in\cite{mirage}.

To complete this argument we must only claim that the Hamiltonian $H_m$ can reproduce field theory dynamics for the limited number of experiments that can be done on the relatively low entropy $m$-block system, over the time scale $N $.  The holographic space-time formalism\cite{hst} makes (incomplete) claims that this can be done within a specific class of Hamiltonians, but one does not need to believe in that formalism to accept the present model.  It is sufficient for {\it some} choice of $H_m$ to "do the right thing".

The model can also describe Hawking radiation that occurs before the scrambling time. The fluctuation that liberates a low energy Hawking particle, described as a decoupled  block of size $\sim 1$, is exponentially unlikely to occur within the still decoupled block $m$, rather than the much larger block $M$.  Thus these particles are not entangled with the infalling system.   If $M \gg m$ this is consistent with the classical computation that shows that almost none of the Hawking radiation comes from the region where a localized quasi-normal mode is ringing down.  Semi-classically we would say that this is a consequence of the fact that Hawking radiation comes out in low angular momentum modes, which have small overlap with a localized perturbation on a large sphere.  

\subsection{No Black Hole Complexity in Linear Dilaton Gravity}

This section will be too short to be very complex.  That is the general message to take away from the comparison between the power law lifetimes of decaying black holes and the exponential time scales for the development of quantum complexity.  In terms of energy scales, complexity deals with the exponentially dense ($e^{- S}\epsilon$) energy spectrum of models with time independent Hamiltonians and many degrees of freedom.  We can model this by replacing the upside down oscillator tails of our Hamiltonian, with a confining potential.  The system then has discrete levels corresponding to the meta-stable states of our discussion, and quantum evolution eventually leads to ergodicity in the space of generic superpositions of these states.  However, in the model we have considered, all of these states acquire a Breit-Wigner width much larger than their splittings, so the dynamics that leads to complexity is irrelevant.

One might try to sustain the discussion of complexity by considering initial states with an influx of particles, at exactly the Hawking rate, which continues for at least the exponential time scales that it takes to grow complexity.  This is considerably easier to do in a model with mass independent Hawking temperature than it is for models with negative black hole specific heat.  However, we are now dealing with an open system and it is not at all clear what we should call "the dimension of the black hole Hilbert space" .   There will always be a subsystem whose nonrelativistic fermion wave functions are concentrated within the range of $K(\lambda, \kappa)$ but that space is entangled with the much larger space of outgoing Hawking radiation states, in a way that varies with time.

\section{Summary and Conclusions}

We have presented a large class of models that can be considered resummations of the perturbation series of $N$ copies of the Type 0B string theory.  One model is completely integrable, and its S-matrix factorizes into that of a tensor product of $N$ copies of Moore's free fermion model.  It has no large entropy meta-stable excitations that can be identified with linear dilaton black holes.  A large class of other models has such excitations but their decay spectrum is not thermal and scattering off these "would be black holes" exactly preserves the $SU(N)$ symmetry of the low energy effective Lagrangian.  An even larger class of models, with of order $N^4/4!$ more couplings than the first, has, for large $N$, meta-stable excitations with all the qualitative properties one expects of black holes.  It is clear that we could produce even more models with these properties if we added higher powers of the fermions to the Hamiltonian, or abandoned some of the rules for the $J_{abcd}$ couplings that make the model soluble at large $N$.  We did not use solubility in our arguments, just the Hartree approximation for $J = 0$ and properties of the SYK model that have been established by numerical work at finite $N$.

We reviewed the work of\cite{cghs}\cite{bddo}\cite{rst1}, which provides a much more complete and seemingly controlled approximation to the quantum field theory in curved space- time approach to this model than recent approaches based on quantum extremal surfaces.   It does not automatically lead to a Page curve, because the solutions of the large $N$ semi-classical equations develop a singularity.  If this singularity is cloaked in a horizon and does not disturb the asymptotic behavior of the fields at null infinity, it would then be a large entropy remnant.  We showed that analysis of the full UV complete model has no such remnant, but that it also has no interpretation of the dynamics inside the horizon in terms of quantum field theory in curved space-time.  In light of this calculation it is not clear how to view attempts to "derive" the Page curve from simple one loop calculations in quantum field theory in curved space-time.  The actual UV complete calculation of the entanglement entropy requires one to understand in detail the subtraction of the vacuum entanglement, which proves to be highly dependent on dynamics of the model that cannot be seen at any order of string perturbation theory.  Failure to do this correctly leads to large errors in the estimate of the entropy of the system in the presence of a black hole.   

Furthermore, the general framework of the island computations of the Page curve would seem to give answers that are relatively independent of modifications of the CGHS Lagrangian, yet there are extant examples where a full semi-classical analysis leads to the conclusion that energy is unbounded from below and black holes never stop decaying. Other examples might exist with large entropy remnants, or with thunderbolts rather than thunderpops.  
These scenarios are inconsistent with the models in this paper and probably with any interpretation in conventional quantum mechanics.

We showed that the island analysis leads to conclusions similar to exact computations in the UV complete model of this paper, so there seems to be something correct about it.  Recent derivations of the island prescription from "replica wormhole" instantons in the bulk Euclidean path integral suggest that it is not a simple one loop approximation.  A version of such calculations has been done for the CGHS model in\cite{hss}. Those arguments should be generalized to study linear dilaton models where black holes do not stop decaying, or which have large entropy remnants or thunderbolts, where the Page curve is the wrong answer.  Alternatively, one could show that no version of asymptotically linear dilaton gravity, has either large entropy remnants or thunderbolts. It's likely that those models are not consistent quantum theories of linear dilaton gravity, but if we are to rely on the island/replica wormhole ideas, they should be able to show that.

Linear dilaton black holes are not a good test case for resolving the firewall problem, because the classical black hole solution does not have a large singularity free area behind the horizon.  In\cite{tb1+1} the author presented a model for the near horizon dynamics of higher dimensional black holes.   This model is very speculative and does not have the many roots in string theory that we have for the linear dilaton models.  Nevertheless, we presented an analysis of the model that showed how the firewall problem {\it could} be resolved.  The basic tenet of the model is that black hole microstates cannot be understood in terms of quantum field theory in curved space-time, but that nonetheless the phenomena uncovered by Hawking are correct coarse grained descriptions of the dynamics, and there is no drama at the horizon for an infalling system.  The model also explains a puzzling feature of black hole entropy formulae for black holes of negative specific heat: the state of matter just before it falls into a black hole has macroscopically smaller entropy than the eventual equilibrium state that develops after the matter has fallen in.  This points to a quantum description in a Hilbert space with many frozen degrees of freedom in the initial state.  If we further postulate that those degrees of freedom must be unfrozen in order to mediate interactions between the matter and the black hole, we see a resolution of the firewall paradox.   The long time of propagation behind the horizon is seen to be the time required to equilibrate the frozen degrees of freedom.  All that is required to make this consistent with an approximate description of the interior by field theory is a Hamiltonian $H_m$ acting on a relatively small number of degrees of freedom, which can mimic field theory behavior for a time scale of order the radius of the black hole.  There is no actual field theory inside, but just a small system, which behaves like field theory until it comes into equilibrium with the non-field theoretic degrees of freedom of the black hole.  This set of ideas leads naturally to models where finite dimensional q-dits are combined into matrices, with Hamiltonians that are single traces of polynomials of those matrices.  

We have come to somewhat ambivalent conclusions about the island prescription.  In its original presentation it seemed to be a one loop correction to classical black hole thermodynamics, and in linear dilaton models it
does not agree with the more sophisticated large $N$ calculations of\cite{cghs}\cite{bddo}\cite{rst1}, unless one chooses models obeying the rules of\cite{rst23}.  Derivation of the island prescription from the gravitational path integral appears to show that it comes from an instanton contribution. It is clear that it cannot capture all of the non-perturbative complexity of actual calculations in the models of this paper, but it does agree with them about the general shape of the Page curve.  This success only exacerbates the
the puzzle of why classical Euclidean computations in general relativity lead one to valid conclusions about the spectrum of quantum theories of gravity. 

The impetus for my return to this problem after $5$ years was an attempt to understand the remarkable calculations of\cite{sss}\cite{sw}, which extract details of the coarse grained energy spectrum of the SYK model from JT gravity path integrals on higher genus surfaces.  I've believed for a long time that our best clue to the connection between classical gravity and quantum spectra is Jacobson's 1995 demonstration\cite{ted} that Einstein's equations were the hydrodynamic equations of the area law for the entropy of the Hilbert space of a causal diamond\cite{bhgtHJFSB} (to use somewhat anachronistic language, and globalize a law that Jacobson used only in infinitestimal form).  Recent derivations of hydrodynamics from quantum theory\cite{hongmukundtbal} reveal that one obtains classical stochastic equations, rather than the classical diffusion or Navier-Stokes equations.  The random "stirring forces" invoked in phenomenological hydrodynamics emerge as part and parcel of the derivation from quantum mechanics, and automatically obey the fluctuation dissipation theorem.  It is well know that classical stochastic equations have a mathematical formulation in terms of functional integrals, which strongly resembles traditional quantum field theory.  In the hydrodynamic context, "loop corrections" to the classical equations are known to lead to long time tails that are missed by classical hydrodynamics.  

To apply this in the context of Jacobson's ideas, one must think about finite causal diamonds, since the holographic principle that engenders these ideas says that most of the states of the quantum theory live on the boundary of the diamond.  Infinite diamonds have infinite entropy and no fluctuations.  Thus one is led to consider classical general relativity in a diamond, with a fluctuating boundary stress tensor. 

\begin{figure}[h!]
\begin{center}
 \includegraphics[width=12cm]{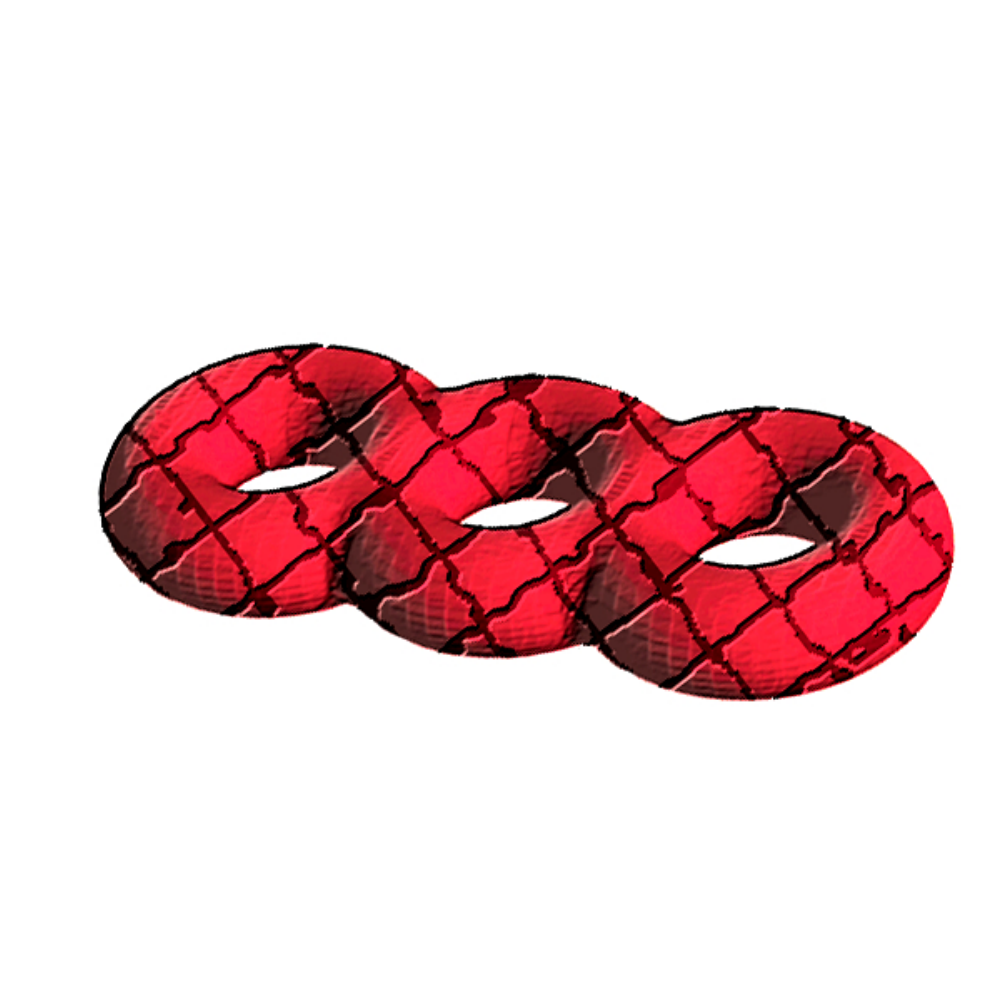}
\end{center}
\vspace{-1.4cm}
\caption{A Geometry Made of Diamonds. Ragged edges are an artist's conception of fluctuations on the boundary.\label{genus3diamond}}
\end{figure}

Equivalently, we can add covariant boundary interaction Lagrangians, multiplied by scalar functions picked from a random distribution.  Following the logic of \cite{Coleman} this should lead to geometries \ref{genus3diamond}.
made of causal diamonds linked together at their boundaries, and plausibly will allow non-trivial topologies.  JT gravity\cite{jt} is a particularly simple example because there is no bulk stress tensor, and the classical equations in the bulk admit only a unique local geometry.   In addition, in two dimensions, the boundary metric of a diamond is purely null and the covariance of the random sources should be independent of the null coordinate.
  There are a lot of details to be filled in but it seems possible that one could rederive the results of\cite{sss} in this way. 

There have been recent suggestions that the results of\cite{sss} point instead to a claim that a path integral over geometries only gives us information about a quenched average over many holographic quantum systems.  Note however\cite{georgesjafferisharlowgurau} that the infinite $N$ SYK model in fact looks exactly like Coleman's wormhole prescription.  That is one performs an average of the partition function over time independent couplings in the boundary theory.  The time in the SYK model is the Schwarzschild time coordinate in $AdS_2$, and can be thought of as a parametrization of the boundary of the maximal causal diamond with purely null boundary.  Perhaps the $t$ independence of the random couplings in the "annealed" approach to SYK is just a reflection of general covariance on a null one dimensional boundary. 
The upshot of these considerations is a conjecture that the sum over topologies in JT gravity is the statistical hydrodynamics of any instantiation of the SYK ensemble. 
I hope to return to an investigation of this conjecture in future work.

The idea that Euclidean path integrals over metrics could be the statistical hydrodynamics of underlying quantum theories might also resolve the puzzle of global symmetries in these $1 + 1$ dimensional models\footnote{In higher dimensional string theory, most symmetries can be shown to be gauged, but some, like dilaton shift symmetry, appear to be global symmetries that we expect to be broken by effects not captured in the perturbation expansion.  The usual argument for this involves black hole physics, but a Euclidean path integral approach in low energy effective field theory would not capture the breaking of these symmetries.}. The Euclidean path integral will preserve all symmetries of the low energy Lagrangian.  This would be puzzling if the path integral were computing S matrix elements, but expected if it were computing only things related to the coarse grained level statistics of the Hamiltonian.

If this conjecture were correct, we would be left with a situation similar to what we've found for linear dilaton gravity.  There are many consistent quantum theories of gravity in $1 + 1$ dimensional space-time with given asymptotic boundaries, but they all lead to the same statistical hydrodynamical equations and the low energy effective Lagrangian is unable to distinguish between different quantum theories. We have not found any criteria that enable us to argue that one or another of the models of quantum linear dilaton gravity that we have constructed is invalid. They are all unitary and have S matrices satisfying standard causality properties, because of the asymptotic connection between $\lambda$ and $r$.  They all have perturbation expansions that agree with string theory.  
In fact, the origin of linear dilaton gravity as the near horizon description of extremal black holes in higher dimension tells us that there are many more consistent models.  The low energy description of those higher dimensional black holes will all have corrections to the effective action in the linear dilaton regime, which are higher derivative terms scaled by the string tension.  The details of these terms will depend
on the compactification manifold on which the ten dimensional theory is reduced to four dimensions and will not agree with stringy corrections to scattering in the $0B$ perturbation expansion.  Calabi-Yau compactifications of Type II string theory are presumably all consistent models, and there should be many models of this type.
In principle each of those models also modifies Moore's fermion Hamiltonian near the top of the inverted oscillator potential, in some particular way, which might or might not be among the choices we exhibited.  

String theorists are used to thinking of models of quantum gravity as more constrained than quantum field theory models.   This is likely to be true in higher dimensions, where evidence points to a connection between exact supersymmetry and low curvature space-times.  We have known for a long time that models in $1 + 1$ dimensions have many fewer constraints.  The models of the present paper simply add to the list.

 \vskip.3in
\begin{center}
{\bf Acknowledgments }\\
\end{center}
This work was supported in part by the U.S. Department of Energy under grant DE-SC0010008.  The author thanks E. Shick for her artistic conception of a genus three surface tiled by fluctuating causal diamonds. He would also like to thank E. Shaghoulian for illuminating email conversations about\cite{hss}, and E.Rabinovici for many suggestions about improving the manuscript.

\section{Appendix A}

The Euclidean  action for our model is
\begin{equation} S = \int dsdtd\lambda d\kappa\ \bigl{[} \psi_a^{\dagger} (t,\lambda) [\partial_t -\partial_{\lambda}^2 - \lambda^2 - \mu \delta (t - s) \delta (\lambda - \kappa)]  + i L(t,\lambda ; s,\kappa ) )\bigr{]}\psi_a (s, \kappa) \end{equation}\begin{equation}+ i  N L(t,\lambda ; s,\kappa )  G(t,\lambda ; s,\kappa ) + [g^2N\delta (t - s) G (t,\lambda ; t, \lambda) G(t,\kappa ; t, \kappa) \end{equation}\begin{equation} + \frac{J_{abcd} }{N^{3/2}} \delta (t - s) \psi_a^{\dagger} (t,\lambda) \psi_b (t,\lambda) \psi_c^{\dagger} (s,\kappa) \psi_d (s,\kappa) ] K(\lambda, \kappa)\bigr{]} . \end{equation}

If we now, following Coleman, average over $J_{abcd}$ with a flavor independent Gaussian distribution, we get a wormhole effective action
\begin{equation} S_{eff} = N\bigl{[} {\rm Tr\ Ln\ } [\partial_t -\partial_{\lambda}^2 -\lambda^2 - \mu + i L(t,\lambda ; s,\kappa) ]\end{equation}\begin{equation} + g^2 \int dt\ d\lambda\ |G|^2 (t,\lambda ; t,\lambda) +  \int dsdtd\lambda d\kappa d\sigma d\rho\ K(\lambda, \kappa ) K(\sigma,\rho) |G(t,\lambda; s,\sigma) |^2 |G(t,\kappa); s,\rho)|^2 .\bigr{]}\end{equation}

Varying this with respect to $G$ and $L$ we get a closed integral equation for the singlet propagator $\frac{1}{N} \langle \psi_a^{\dagger} (t,\lambda) \psi_a ( s, \kappa) \rangle . $   This equation has no sign of $SU(N)$ breaking.  The Colemanesque wormhole averaging is equivalent to quenched averaging in the large $N$ limit\cite{georgesjafferisharlowgurau}.

\section{Appendix B}

The contents of this appendix summarizes part of an unpublished manuscript by the author, W. Fischler, and Daniel Park.  

Our system consists of non-relativistic fermion fields in an inverted oscillator potential cutoff by walls in $\lambda$ space.  The individual oscillator modes have a discrete spectrum, and each of them gives us a Hilbert space of fermionic oscillators transforming in the $[2] \oplus ... \oplus [2N]$ representation of $SU(2)$.  We compute 
\begin{equation} e^{W(\nu)} \equiv {\rm Tr} e^{ - \nu J_3} = \prod_l (\sum_{m = - l}^{l}(1 + e^{-\nu m}) = e^{\sum_l {\rm ln}\ [2\sum_{m = 1/2}^l(1 + \cosh (\nu m))] }. \end{equation}  
For large $N$ this is approximately 
\begin{equation} W(\nu) = \int_0^N dl\ {\rm ln}[2\int_0^l dm\ ( 1 + \cosh (\nu m))] . \end{equation} Rescaling both integration variables by $N$ this becomes
\begin{equation} W(\nu) = N \int_0^1 dl\ {\rm ln}[2N\int_0^l dm\ (1 + \cosh (\nu N m))] . \end{equation} 

On the other hand
\begin{equation} Z(\nu) = \sum_{l = 1/2}^{N - 1/2} e^{S(l)} \sum_{m = 1/2}^l \cosh (\nu m) , \end{equation} which is approximately
\begin{equation} Z(\nu) =N^2 \int_{0}^{1} dl\ e^{S(l)} \int_{0}^l dm \cosh (\nu N m) , \end{equation} where $e^{S(l)}$ is the number of states in the Hilbert space with spin $l$.   Our exact calculation of the large $N$ limit of $Z(\nu)$ shows that the distribution $S(l)$ is gaussian, with a width of order $N^2$, despite the fact that angular momenta as large as $N^3$ are allowed in our Hilbert space.   We added the term $G({\bf L}^2)$ to the Hamiltonian of our black hole model in the text, in order to restrict the width and suppress higher angular momenta more than the infinite temperature ensemble does.  A more ambitious goal would be to choose $G$ in order to reproduce the rotating black hole entropy formula for $S(l)$.  This can be done but the answer did not seem terribly illuminating, which accounts for the fact that no manuscript of this work was published.

\section{Appendix C}

Consider matrix quantum mechanics with Euclidean action
\begin{equation} S = N^2  \int dt\  {\rm Tr\ } (\dot{M}/N)^2  + V(M/N)) , \end{equation} where the $U(N)$ invariance is treated as a gauge symmetry.  In Hilbert space terms this means that we restrict attention to the singlet sector.  This model has a standard $1/N$ expansion where Feyman graphs of genus $g$ come with a power $N^{2 - 2g}$. The graphs depend on the $N$ independent 't Hooft couplings encoded in the polynomial $V$.  

If we integrate over the unitaries that diagonalize the matrix $M$, we obtain a completely gauge invariant description in terms of its eigenvalues.  The residual $S_N$ gauge symmetry makes the eigenvalues into non-relativistic Bose particles.  The action is
\begin{equation} S = N [\sum \dot{\lambda_i}^2 + V(\lambda_i) + {\rm ln\ } \sum_{i < j} (\lambda_i - \lambda_j)^2 ] . \end{equation} 
We can absorb the interaction into the integration measure of the $\lambda_i $  by agreeing to multiply the initial and final state wave functions by $\prod_{i < j} (\lambda_i - \lambda_j)$.  This function is odd under interchange of any two eigenvalues and turns the eigenvalues into fermions.

In the large $N$ limit the eigenvalue spectrum becomes continuous.  If we simultaneous focus on a maximum of the potential, and tune the fermi surface to be near the maximum, then we obtain a limit where the sum over discrete genus $g$ surfaces is replaced by an integral over continuous world sheets, with a conformal field theory (including Fadeev-Popov ghosts) of central charge $0$ propagating on them.  
In the limiting theory, there are two parameters, a scale defined by the curvature of the potential, and the distance of the fermi surface below the maximum, in units of that curvature.  The latter parameter controls the genus expansion.

The world sheet theory has two scalar fields, one a Liouville field with anomalous conformal transformation properties, and the other an ordinary scalar, whose origin is the $\lambda$ coordinate.  The result is interpreted as string theory in two space-time dimensions.  The anomalous transformation law of the Liouville field fixed the dilaton field to be a linear function of the target space coordinate, in the frame where the target space metric is Minkowski.  This means that The other target space fields are two tachyons (these fields are actually massless in the asymptotic region, as a consequence of the non-trivial dilaton profile, but the name tachyon has been carried over from string theory in higher dimensions).  The original formulation of the theory had only one of these tachyons and corresponds to bosonic string theory.  The fermi surface depth is interpreted in the world sheet theory as the inverse string coupling.  Lowering the fermi surface below the top of the potential is equivalent, in the world sheet theory to adding a tachyon background on target space.   This acts as a potential barrier for scattering of tachyons, which prevents them from entering the region of space-time where the string coupling is strong.  From the point of view of non-relativistic fermions, the low fermi surface prevents the fermions from reaching the maximum of the potential.

It's interesting, and a bit sad, that a deep insight about the meaning of strong string coupling, which was appreciated in at least some of the work on the CGHS model, was completely missed by those who worked on the matrix model/fermionic/string theoretic formulation of these models\footnote{The present author, who worked on both of these topics, is more guilty than most for missing this connection.  He didn't understand it until\cite{tb1+1} in 2015.}.   {\it From the point of view of semi-classical gravity, strong string coupling is synonymous with low entropy.}  The space-like singularity inside a classical linear dilaton black hole is exactly a point of zero entropy.   

We can rederive the gravitational picture of the distribution of entropy in space-time directly from the non-relativistic fermionic field theory.  In the Thomas-Fermi approximation, the entropy stored in the region $[0,\lambda]$ in the fermion ground state at $\mu = 0$ is
\begin{equation} \int dp\ \theta (- p^2 + \lambda^2) \propto \lambda . \end{equation} As we'll review in a moment, the space-time coordinate far out along the $\lambda$ axis at fixed $p$ is ${\rm ln}\ \lambda$, so the entropy grows exponentially with $\lambda$, which is the linear dilaton profile.

One can have different fermi surface levels on the two sides of the origin, but in the bosonic version of the theory the "other side" is taken to be the empty fermi sea and the model is unstable.  The instability is invisible to all orders in string perturbation theory.   The stable theory has two tachyon eigenstates corresponding to wave functions that are even or odd under reflection of $\lambda$.  Because the Liouville theory is an interacting, though integrable, quantum field theory, string perturbation computations are highly non-trivial, and only recently have they been able to confirm the results of the non-perturbative S-matrix we'll explain below\cite{yinetal}.  Even these low order results required numerical integration.

The S matrix is most efficiently calculated by going to the variables $x_{pm}$.  The wave function in one of these variables is just the Fourier transform of that in the other basis.  On the other hand, going to the variables $r_{\pm} = {\rm ln} (x_{\pm})$ the Hamiltonian (with $\mu = 0$) just becomes \begin{equation} H = \pm \frac{\partial}{\partial r_{\pm}} , \end{equation} which is the Hamiltonian of left and right moving Weyl fermions.   

Written in terms of the coordinates $x_{\pm}$ the eigenstates of the Hamiltonian are singular at $\lambda = 0$, and we have to treat the positive and negative halves of the $\lambda$ axis as two different pairs of incoming/outgoing Weyl fermions. With these clues, and consulting\cite{akk} for details, the reader should be able to construct the exact S-matrix of the 0B model.

\end{document}